\documentclass[runningheads]{llncs}

\usepackage[T1]{fontenc}
\usepackage{graphicx}
\usepackage{booktabs}
\usepackage[misc]{ifsym}

\usepackage{amsmath,amsfonts}
\usepackage{algorithmic}
\usepackage{makecell}
\usepackage{xcolor}
\usepackage{todonotes}
\usepackage{multirow}
\usepackage{threeparttable}
\usepackage{enumitem}
\usepackage{colortbl}

\begin{document}

\title{The Hyperscale Lottery: How State-Space Models Have Sacrificed Edge Efficiency}
\titlerunning{The Hyperscale Lottery}

\author{{Robin Geens\inst{1}$^,$\inst{2} \and Jonas De Schouwer\inst{2} \and Marian Verhelst\inst{1} \and Thierry Tambe\inst{2}}}

\authorrunning{R. Geens et al.}

\institute{MICAS, KU Leuven, Leuven, Belgium
\and Stanford University, Stanford, CA, USA \\
\email{robin.geens@kuleuven.be}
}

\maketitle

\begin{abstract}

The Hardware Lottery posits that research directions are dictated by available silicon compute platforms. We identify a derivative phenomenon, the Hyperscale Lottery, where model architectures are optimized for cloud throughput at the expense of algorithmic efficiency. While State-Space Models (SSMs) such as Mamba were lauded for their linear complexity---ideal for edge intelligence---their evolution from Mamba-1 to Mamba-3 reveals a systematic divergence from edge-native efficiency. We demonstrate that Mamba-3’s architectural changes, designed to saturate hyperscale GPUs, impose a significant edge penalty: a 48\% latency increase for 15M-parameter edge models, with the penalty persisting at 22\% even at 880M. We argue for decoupling cloud-scale saturation strategies from core architectural design to preserve the viability of single-user, real-time edge intelligence.

\end{abstract}

\section{Introduction}

\begin{figure}
    \centering
    \includegraphics[width=0.9\linewidth]{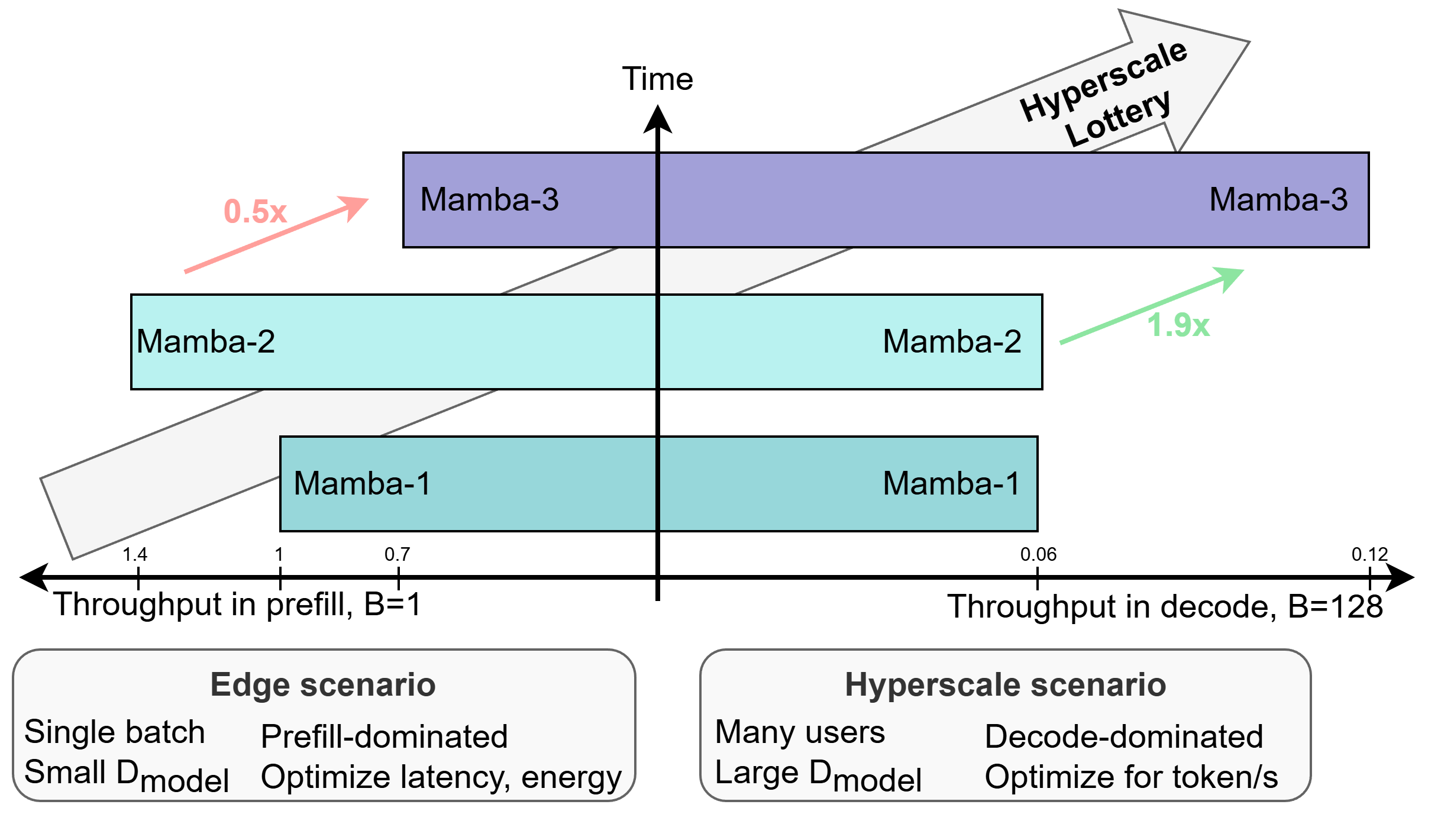}
    \caption{Normalized throughput of the state-update of different Mamba architectures  on a hypothetical edge accelerator, estimated with a roofline model. \\ 
    }
    \label{fig:roofline}
\end{figure}

While the Hardware Lottery~\cite{hw-lottery} dictates that available silicon shapes Machine Learning research, we observe a derivative phenomenon. To maximize impact, new architectures are born in the cloud, designed to maximize aggregate throughput by hiding memory latency through massive batch sizes ($B \gg 1$) and multi-GPU distributed strategies such as Pipeline Parallelism and Tensor Parallelism. This leads to the Hyperscale Lottery: architectures evolve toward compatibility with economic and operational imperatives of hyperscalers, rather than advancing the theoretical Pareto front of accuracy versus raw computational complexity.

This shift  represents a systematic departure from edge-native requirements. While cloud environments prioritize aggregate throughput ($\text{tokens}/s$), edge applications (e.g., physical AI, AR/VR, privacy-preserving LLMs) demand low latency ($ms/token$), energy efficiency, and high performance at a batch size of $B=1$. By baking cloud-scale saturation strategies directly into model topologies, the current research trajectory leaves improvements in real-time, privacy-preserving, and energy-constrained intelligence off the table.

To ground this claim quantitatively, we (i) re-derive per-token operation counts for Mamba-1, -2, and -3 from a unified analytical formulation; (ii) extend an open-source modeling framework to support Mamba-3's MIMO state expansion on edge ASIC architectures; and (iii) compare prefill latency and energy across model sizes using GPU measurements, analytical modeling, and roofline analysis.

\section{SSMs: the promise for edge intelligence}

State-Space Models (SSMs) have emerged as the primary alternative to Transformers, offering $\mathcal{O}(L)$ computational complexity against attention's $\mathcal{O}(L^2)$. This linear efficiency becomes advantageous for sequence lengths $L > 6 D_{\text{model}}$ \cite{mambaout}, with $D_{\text{model}}$ being the embedding dimension. For edge intelligence, this threshold is particularly significant: edge-deployed models typically operate with a smaller $D_{\text{model}}$, which means SSMs become arithmetically superior even at shorter sequence lengths, including those encountered in vision tasks. Furthermore, the fixed-size recurrent state ensures a constant memory footprint, a crucial requirement for hardware with strict memory constraints and limited dynamic allocation capabilities.

This efficiency has driven rapid adoption in computer vision~\cite{vision-mamba,vmamba,videomamba,simba}, physical AI~\cite{robomamba,drama}, and Small Language Models~\cite{hymba,nemotron,zamba}. Although Mamba was initially proposed as an LLM backbone, its success in these edge-centric applications underscores its role as a versatile architecture for resource-constrained intelligence.

\section{The Mamba Mutations}

Due to its popularity, the Mamba family has emerged as the \textit{de facto} representative of SSMs. However, the architecture has undergone drastic structural changes during its development, revealing a clear trajectory toward hyperscale optimization.

\textbf{Mamba-1}~\cite{mamba1} relies on a selective scan mechanism. While the algorithmic formulation is purely sequential (i.e., every timestep in $L$ is processed in order), the original paper also proposes a parallel scan (pscan) implementation to increase hardware utilization on GPUs at the cost of more operations. However, pscan does not alter the fundamental recurrent dynamics of the model and is effectively a deployment-time kernel optimization, preserving Mamba's edge-native suitability.

\textbf{Mamba-2}~\cite{mamba2} marks the inflection point toward cloud compatibility through two distinct architectural pivots. First, to enable the Structured State Space Duality (SSD) formulation, which allows the model to be computed via matrix multiplications on Tensor Cores, the state transition matrix $A$ is restricted from a diagonal matrix to a scalar structure. This simplification results in the loss of per-channel decay, sacrificing fine-grained temporal expressivity for the sake of compute-bound hardware utilization. 
Second, Mamba-2 introduces a "head" dimension distinct from the channel dimension. This structural change enables Tensor Parallelism across multiple GPUs, allowing for a single all-reduce synchronization step per layer. Ultimately, these modifications successfully align the architecture with Transformer-based distributed inference infrastructure and hyperscale deployment practices.

\textbf{Mamba-3}~\cite{mamba3} introduces the most significant edge penalty yet by incorporating a MIMO rank dimension ($R=4$) into the state expansion. This converts state updates from vector-based outer products into matrix-matrix multiplications, increasing the Operational Intensity (OI) by $R\times$. The motivation is clear in high-batch decode regimes: when the state update is memory-bound, a higher OI enables better utilization of compute resources, yielding a theoretical $R\times$ speedup. However, this benefit is insignificant for single-batch ($B=1$) decode and actively harmful during the prefill phase, where the additional computation provides no throughput benefit whatsoever. To maintain a constant parameter count, the model dimension $D_{\text{model}}$ is scaled by $1/\sqrt{R} = 0.5$. Because SSM operations scale as $L \cdot D \cdot N \cdot R$, the net effect is a $2\times$ increase in computation and energy consumption per token. Although the Mamba-3 paper also describes a SISO variant, it treats it only as an ablation while positioning MIMO as the headline architectural advance.

These evolutionary steps highlight a consistent pattern: modern SSM development prioritizes high-throughput saturation on hyperscale GPUs at the direct expense of the per-token efficiency that originally made Mamba-1 a compelling candidate for edge deployment.

\begin{table}[t]
\caption{Performance of Mamba variants on an 880M-parameter
         model with $B=1$. Indicated changes are relative to Mamba-2.}
\label{tab:results}
\centering
\begin{threeparttable}
\small
\setlength{\tabcolsep}{5.5pt}
\begin{tabular}{l cc cc cc}
\toprule
\textbf{}
  & \multicolumn{2}{c}{\textbf{Mamba-1}}
  & \multicolumn{2}{c}{\textbf{Mamba-2}}
  & \multicolumn{2}{c}{\textbf{Mamba-3}} \\
\cmidrule(lr){2-3}\cmidrule(lr){4-5}\cmidrule(lr){6-7}
  & \textbf{Seq.} & \textbf{pscan}
  & \textbf{Seq.} & \textbf{SSD}
  & \textbf{Seq.} & \textbf{SSD} \\
\midrule
\rowcolor{gray!20}
\multicolumn{7}{l}{\textbf{Operations}} \\
Total [GOps/token]
  & 1.52 & 1.56 & 1.43 & 1.46
  & 1.62 {\color[HTML]{FE0000}\scriptsize(+13\%)} & 1.71 \\
State-update\tnote{a}\ $\;$ [GOps/token]
  & 0.066 & 0.104 & 0.048 & 0.075
  & 0.098 {\color[HTML]{FE0000}\scriptsize(+100\%)} & 0.189 \\
Op. Intensity\tnote{a} $\;$ [ops/Byte]
  & 53.2 & 76.7 & 50.8 & 31.7 & 49.2 & 37.0 \\
\midrule
\rowcolor{gray!20}
\multicolumn{7}{l}{\textbf{A100 (measured)\tnote{a}}} \\
Throughput [tok/s]
  & 585 & 18\,002 & 735 & 22\,601
  & \multicolumn{2}{c}{N/A\tnote{b}} \\
Peak mem.\ [MB]
  & 54.0 & 86.3 & 55.5 & 90.4
  & \multicolumn{2}{c}{N/A\tnote{b}} \\
\midrule
\rowcolor{gray!20}
\multicolumn{7}{l}{\textbf{Edge ASIC (Stream)}} \\
Throughput [tok/s]
  & 549 & N/A\tnote{c} & 602 & N/A\tnote{c}
  & 495 {\color[HTML]{FE0000}\scriptsize($-$18\%)} & N/A\tnote{c} \\
Energy [mJ/tok]
  & 5.85 & N/A\tnote{c} & 5.37 & N/A\tnote{c}
  & 6.86 {\color[HTML]{FE0000}\scriptsize(+28\%)} & N/A\tnote{c} \\
\midrule
\rowcolor{gray!20}
\multicolumn{7}{l}{\textbf{Edge ASIC (Roofline)}} \\
Throughput [tok/s]
  & 673 & 657 & 714 & 700 & 634 \\
\bottomrule
\end{tabular}
\begin{tablenotes}[para]
  \scriptsize
  \item[a] State-update only, all layers. \\
  \item[b] GPU kernel not available open-source at the time of writing. \\
  \item[c] Stream does not support SSD formulations.
\end{tablenotes}
\end{threeparttable}
\end{table}

\section{Experiments}

To quantify the \textit{edge penalty} imposed by the hyperscale-optimized architectures, we evaluate the inference characteristics of all Mamba variants using three distinct methodologies:

\begin{enumerate}[topsep=2pt, partopsep=0pt, parsep=2pt, itemsep=1pt, leftmargin=*]
    \item \textbf{GPU baseline}. We run the open-source GPU kernels on a single NVIDIA A100 GPU to establish the cloud-native performance ceiling.
    \item \textbf{Analytical modeling}. To estimate performance on edge chips, we use the Stream modeling framework~\cite{stream}, which supports a variety of hardware architectures and has explicit support for compute-bound, sequential SSM formulations~\cite{fine-grained}. We extend Stream with support for Mamba-3's MIMO state expansion and RoPE embeddings.
    \item \textbf{Roofline analysis}. We apply a roofline model to determine the maximal theoretical throughput based on the operational intensity of each operator.
\end{enumerate}

\noindent For the analytical and roofline models, we define a hardware architecture representative of modern edge accelerators: a MAC array of 1024 processing elements and a SIMD array of 32 lanes operating at 500 MHz, supported by 2 MB of on-chip SRAM and two channels of LPDDR5, providing a total of 34 GB/s off-chip bandwidth. We assume an average cost of 15 pJ/bit of off-chip memory traffic~\cite{mem-energy} and an aggregated 2 pJ/op for computation~\cite{op-energy}.

The results in Table~\ref{tab:results} demonstrate that sequential formulations remain the optimal choice for Edge ASICs across all variants. The transition to Mamba-2 initially yields a modest improvement in edge throughput, attributable to the algorithmic simplification of the state transition matrix reducing the total number of state-update operations. The full impact of the Hyperscale Lottery materializes in Mamba-3. Engineered to maximize decode throughput in high-batch cloud environments, Mamba-3 exhibits a clear regression in the prefill phase: a $2\times$ increase in state-update operations results in a $-18\%$ throughput ($+22\%$ latency) penalty relative to Mamba-2. 

\begin{figure}
    \centering
    \includegraphics[width=0.9\linewidth]{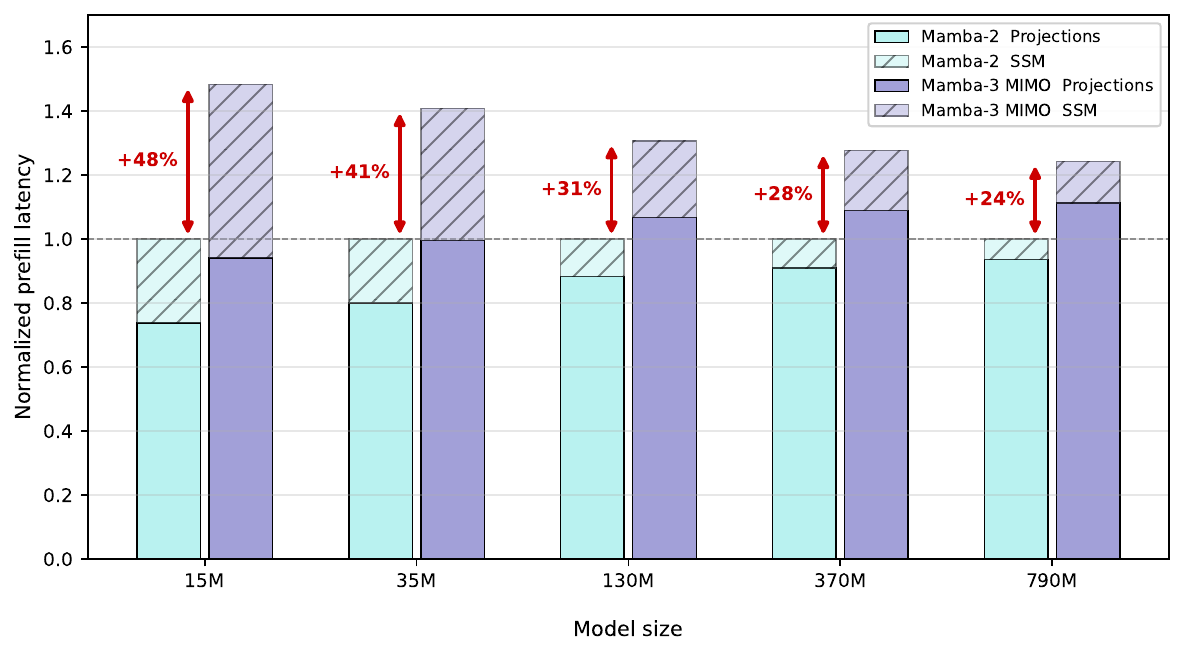}
    \vspace{-4mm}
    \caption{Normalized latency of different model sizes of Mamba-2 and Mamba-3 MIMO, estimated with a roofline model.}
    \label{fig:scaling}
\end{figure}

Figure~\ref{fig:roofline} visualizes these diverging priorities. In the left panel (edge prefill, $B=1$), throughput decreases from Mamba-1 to Mamba-3. In the right panel (hyperscale decode, $B \gg 1$), the trajectory reverses: Mamba-3's higher operational intensity pays off. The architectural evolution from Mamba-1 to Mamba-3 represents a re-targeting toward a different deployment regime rather than a uniform improvement. While edge batch sizes may grow beyond $B=1$ in agentic or multi-application scenarios, $B=1$ remains the critical design point for latency-sensitive applications such as physical AI and AR/VR, where batching improves aggregate throughput but not per-request latency.

Crucially, the $+22\%$ latency penalty observed at 880M parameters is not static: it worsens as models shrink. Dense projection operations scale quadratically with model dimension ($\propto D_{\text{model}}^2$), while state-update operations scale only linearly ($\propto D_{\text{model}}$). Shrinking a model for edge deployment therefore disproportionately increases the relative contribution of state-update latency to total inference time, amplifying the cost of Mamba-3's MIMO overhead. As illustrated in Figure~\ref{fig:scaling}, the latency penalty increases to $+48\%$ for a 15M-parameter model.

\section{Conclusion}

The Hyperscale Lottery reveals a structural tension in contemporary SSM research: the modifications that make Mamba competitive with Transformers at cloud scale are precisely the modifications that degrade its performance at the edge. By embedding cloud-scale saturation strategies directly into the model architecture, the newest SSM architecture forfeits the deployment flexibility that a purely algorithmic optimization would preserve.

To prevent a monoculture of cloud-exclusive AI architectures, future hardware-algorithm co-design must explicitly branch to accommodate the strict latency, memory, and energy constraints of single-batch edge deployments. Concretely, this means treating batch-size-dependent optimizations as deployment-time transforms rather than training-time architectural constraints, and evaluating new architectures against edge benchmarks alongside the aggregate throughput metrics that currently dominate the literature.

\begin{credits}
\subsubsection{\ackname}
This project has been partly funded by the European Research Council (ERC) under grant agreement No. 101088865, the European Union’s Horizon 2020 program under grant agreement No. 101070374, the Flanders AI Research Program, Research Foundation Flanders (FWO) under grant No. 1S37125N, KU Leuven, and Stanford University.
\end{credits}

\bibliographystyle{splncs04}
\bibliography{references_short}

\end{document}